\documentclass[twocolumn]{jpsj2} 
\input epsf.tex  

\title{Thermodynamic Relations in Correlated Systems}

\author{Shinji \textsc{Watanabe} and Masatoshi \textsc{Imada}}

\inst{Institute for Solid State Physics, 
University of Tokyo, 5-1-5, Kashiwanoha, Kashiwa, Chiba 277-8581
}

\abst{
Several useful thermodynamic relations are derived for metal-insulator
transitions,
as generalizations of the Clausius-Clapeyron and Eherenfest theorems.
These relations hold in any spatial dimensions and at any temperatures.
First, they relate several thermodynamic quantities to the slope of the
metal-insulator
phase boundary drawn in the plane of the chemical potential and
the Coulomb interaction in the phase diagram of the Hubbard model.
The relations impose constraints on the critical properties of the Mott
transition.
These thermodynamic relations are indeed confirmed to be satisfied
in the cases of the one- and two-dimensional Hubbard models.
One of these relations yields that
at the continuous Mott transition with a diverging charge compressibility,
the doublon susceptibility also diverges.
The constraints on the shapes of the phase boundary
containing a first-order metal-insulator transition at finite temperatures
are clarified based on
the thermodynamic relations. For example, the first-order phase boundary 
is parallel to the temperature axis asymptotically 
in the zero temperature limit. 
The applicability of the thermodynamic relations are not restricted only to
the metal-insulator transition of the Hubbard model,
but also hold in correlated systems with any types of phases in general.
We demonstrate such examples in an extended Hubbard model with intersite
Coulomb repulsion containing the charge order phase.
}

\kword{charge compressibility, doublon susceptibility, Mott transition, thermodynamic relation}

\begin{document}
\maketitle

\section{Introduction}
The interplay between the kinetic energy and the correlation effect for 
Fermion systems has been one of the most important subjects in the condensed matter physics. 
The interplay of quantum fluctuations and many-body effects can induce highly non-trivial phenomena. 
The metal-insulator transition driven by the electron correlation, which is called the 
Mott transition~\cite{Mott}, offers such a prototype~\cite{IFT}. 

In the Mott transition, there are two control parameters: 
one is the chemical potential and the other is the bandwidth defined by the relative 
strength of the Coulomb interaction to the hopping integral. 
Controls by these two parameters are realized in a lot of examples 
in real materials as 
transition-metal compounds~\cite{IFT} including cuprates which exhibits 
the high-$T_{c}$ superconductivity, 
organic materials~\cite{Organic} and $^3$He systems~\cite{He}. 

Recently, critical exponents of the Mott transition 
through the bandwidth-control route 
in the phase diagram of temperature, pressure and magnetic field 
have been reported in (V$_{1-x}$Cr$_x$)$_2$O$_3$~\cite{LGJWMH}. 
The critical point was also identified 
in $\kappa$-(ET)$_2$Cu[N(CN)$_2$]Cl~\cite{KIMK}. 
Although 
the measurements in (V$_{1-x}$Cr$_x$)$_2$O$_3$ support 
the Ising universality~\cite{LGJWMH} 
predicted by G. Kotliar, {\it et al}.~\cite{Kotliar1,RCK,KLR}
within the dynamical-mean-field-theory
in the infinite spatial dimension~\cite{GKKR}, 
the measurements in $\kappa$-(ET)$_2$Cu[N(CN)$_2$]Cl suggest a new aspect for 
the Mott transition~\cite{KIMK}. 
These circumstances indicate 
the importance of the theoretical study 
to clarify the basic properties of the Mott transition and 
the mutual relationship among the basic physical quantities  
near the Mott transition in finite spatial dimensions. 


Theoretical studies of the Hubbard model by fully taking into account the spatial 
fluctuations in finite dimensions have been done by numerical calculations 
and analytical methods. 
For the filling-control Mott transition, 
the continuous character with the diverging charge compressibility 
was clarified at zero temperature in the one-dimensional Hubbard model 
by the Bethe-Ansatz calculation~\cite{UKO} and in the two-dimensional Hubbard model 
by the quantum Monte Calro (QMC) calculations~\cite{FI,FI2}. 

For the bandwidth-control Mott transition, the ground-state phase diagrams 
of the half-filled Hubbard models on the square lattice~\cite{KI} and 
the anisotropic triangular lattice~\cite{MWI} have been determined by 
the path-integral renormalization group (PIRG) method~\cite{PIRG1,PIRG2}. 
It has been suggested that the order of the Mott transitions changes from the 
first-order to continuous ones as the effect of magnetic frustration becomes large. 

Recently, a new numerical algorithm, the grand-canonical path-integral renormalization 
group (GPIRG) method has been developed~\cite{WI}, 
which enables us to study the filling- and bandwidth-control Mott transitions 
in a unified framework. 
The GPIRG method has first made it possible to determine the ground-state phase diagram in the plane of 
the chemical potential and the Coulomb interaction of the square-lattice Hubbard model~\cite{WI}. 
It has been found that the V-shaped Mott insulator phase appears with the first-order 
bandwidth-control Mott transition at the corner of the V shape and 
the continuous filling-control Mott transition at the edges.  
The V-shaped structure 
has been shown to be consistent with the contrast found in the difference 
between the first-order bandwidth- and the continuous filling-control routes. 
This consistency is derived from the generalized Clausius-Clapeyron and Ehrenfest relations 
between the slope of the metal-insulator boundary and physical quantities~\cite{WI}. 

In this paper, we further study the thermodynamic relations. 
We show that a couple of general relations between physical quantities 
hold at each case of the first-order and continuous Mott transitions. 
These relations are again derived from the Clausius-Clapeyron 
and Ehrenfest relations and 
hold at any spatial dimensions and at any temperatures. 
We derive the thermodynamic relations between the charge compressibility 
and the doublon susceptibility for the continuous Mott transition, 
where the doublon susceptibility is shown to diverge when the charge compressibility diverges.
The relations impose constraints on the shapes of phase diagrams. 


The organization of this paper is as follows:
In \S2, the basic physical quantities for filling- and bandwidth-control routes are defined.
The thermodynamic relations between the slope of the phase boundary and the physical 
quantities are derived for each case of the first-order and continuous Mott transitions. 
In \S3, 
it is shown that they are actually satisfied in analytic forms obtained 
in the one- and two-dimensional Hubbard models.  
In \S4, the thermodynamic relations are extended to finite temperatures. 
The constraints on the shapes of the phase diagram for the finite-temperature 
first-order transition are classified by using the thermodynamic relations. 
In \S5, an application to other systems is demonstrated 
by taking an example of an extended Hubbard model with intersite Coulomb repulsion. 
We summarize the paper in \S5.

\section{Derivation of thermodynamic relations} 

\subsection{Basic physical quantities for filling- and bandwidth-control routes}
The Hubbard model which we consider is 
\begin{eqnarray}
H&=&-\sum_{{\langle i,j \rangle}\sigma}
t_{ij}
\left(
c^{\dagger}_{i\sigma}c_{j\sigma} + c^{\dagger}_{j\sigma}c_{i\sigma}
\right)
-\mu \sum_{i\sigma}n_{i\sigma}
\nonumber \\
& &+U\sum_{i=1}^{N}
\left( n_{i\uparrow} -\frac{1}{2}\right)
\left( n_{i\downarrow}-\frac{1}{2}\right),         
\label{eq:Hamil}
\end{eqnarray}
where $c_{i\sigma}$ $(c^{\dagger}_{i\sigma})$
is the annihilation (creation) operator on the $i$-th site 
with spin $\sigma$ and $n_{i\sigma}=c^{\dagger}_{i\sigma}c_{i\sigma}$
in the $N$-lattice system. 
Here, $t_{ij}$ is the transfer integral and in this paper, the nearest-neighbor transfer 
is taken as the energy unit. 
In eq.~(\ref{eq:Hamil}) 
$\mu$ is the chemical potential and $U$ is the Coulomb interaction. 

The electron filling $n$ and the double occupancy $D$ are defined by the first derivative of the Hamiltonian as 
\begin{eqnarray}
n &\equiv& -\frac{1}{N} 
\frac{\partial \langle H \rangle}{\partial \mu}
= \frac{1}{N}
\sum_{i=1\sigma}^{N}\langle n_{i\sigma} \rangle, 
\label{eq:filling}
\\
D &\equiv& \frac{1}{N}
\frac{\partial \langle H \rangle}{\partial U}
\nonumber
\\
&=& \frac{1}{N}
\sum_{i=1}^{N}\left\langle \left(
n_{i\uparrow}-\frac{1}{2}
\right)
\left(
n_{i\downarrow}-\frac{1}{2}
\right)\right\rangle, 
\label{eq:dbleoccp}
\end{eqnarray}
where $\langle \cdots \rangle$ represents the ground-state expectation value or 
the thermal average at finite temperature. 

The charge compressibility $\chi_{c}$ and the doublon susceptibility $\chi_D$ are defined, 
respectively, by the second derivative of 
the Hamiltonian as
\begin{eqnarray}
\chi_{c} &\equiv& 
-\frac{1}{N} 
\frac{\partial^{2} \langle H \rangle}{\partial \mu^{2}}
= \left(
\frac{\partial n}{\partial \mu}
\right),
\label{eq:chargecomp}
\\
\chi_{D}&\equiv& 
-\frac{1}{N}
\frac{\partial^{2} \langle H \rangle}{\partial U^{2}} 
= -\left(
\frac{\partial D}{\partial U}
\right).
\label{eq:doblonsus}
\end{eqnarray}
%

\subsection{The case where the phase boundary $U(\mu)$ has finite slope $\delta U/\delta \mu$}
\subsubsection{First-order-transition case}
Although the following derivation of eq.~(\ref{eq:derivum}) 
has been given in ref.~\cite{WI}, we show here it 
for the self-contained description. 
Let us consider the ground state, 
although the results described below hold at finite temperatures, 
as will be mentioned in \S4. 
The ground-state energy is expanded by 
$\mu$ and $U$: 
\begin{eqnarray}
\lefteqn{E(\mu+\delta\mu,U+{\delta}U)}\nonumber \\ 
&=&E(\mu,U)+\left(\frac{{\partial}E}{{\partial}\mu}\right)_{U  }\delta\mu +\left(\frac{{\partial}E}{{\partial}U  }\right)_{\mu}{\delta}U \nonumber \\
& &+O(\delta\mu^2,\delta U^2). 
\label{eq:E1}
\end{eqnarray}
In the $\mu$-$U$ phase diagram, along 
the metal-insulator-transition boundary, 
the energies of the metal and insulator phases should be the same because of the coexistence: 
\begin{eqnarray}
\lefteqn{E_{I}(\mu,U)-E_{M}(\mu,U)}\nonumber \\
& & = E_{I}(\mu+\delta\mu,U+{\delta}U) - E_{M}(\mu+\delta\mu,U+{\delta}U)\nonumber \\
& & = 0, \label{eq:E2}
\end{eqnarray}
where subscripts I(M) represent that the expectation value 
is taken in the insulating (metallic) ground state. 
If the first-order metal-insulator transition occurs, 
the slope of the transition line in the $\mu$-$U$ phase diagram is 
determined by the ratio of the jump of filling and double occupancy. 
Namely, from eq.~(\ref{eq:E1}) and eq.~(\ref{eq:E2}) 
the following equation is derived: 
\begin{eqnarray}
\frac{{\delta}U}{{\delta}\mu}=\frac{n_{I}-n_{M}}{D_{I}-D_{M}}. 
\label{eq:derivum}
\end{eqnarray}

\subsubsection{Continuous-transition case}
Let us consider the phase diagram in the $\mu$-$U$ plane in the ground state, $T=0$. 
If the continuous metal-insulator transition occurs and the slope of the transition line is finite 
in the $\mu$-$U$ plane, the double occupancy is expanded by $\mu$ and $U$, as follows:
\begin{eqnarray}
D(\mu+\delta\mu, U+\delta U)=D(\mu,U) & &
\nonumber
\\
+\left(\frac{\partial D}{\partial \mu}\right)_{U}\delta\mu
                      +\left(\frac{\partial D}{\partial U}\right)_{\mu}\delta U
                       &+&O({\delta \mu}^2,{\delta U}^2).
\label{eq:Dexpand}
\end{eqnarray}
Along the metal-insulator-transition boundary, 
the double occupancies at the metallic and insulating phases 
have the same value at $(\mu,U)$ and $(\mu+\delta\mu,U+\delta U)$: 
\begin{eqnarray}
D_{M}(\mu+\delta\mu,U+\delta U)-D_{I}(\mu+\delta\mu,U+\delta U) & &
\nonumber 
\\
=D_{M}(\mu,U)-D_{I}(\mu,U),& &
\label{eq:Deqs}
\end{eqnarray}
where the subscripts M(I) represent that the average 
$\langle \cdots \rangle$ in eq.~(\ref{eq:dbleoccp}) 
is calculated in the metallic (insulating) state. 
By substituting eq.~(\ref{eq:Dexpand}) to eq.~(\ref{eq:Deqs}), the slope of 
the metal-insulator-transition boundary is expressed as 
\begin{eqnarray}
\frac{\delta U}{\delta \mu}&=&-\frac{\left.\left(\frac{\partial D}{\partial \mu}\right)_{U}\right|_{M}
-\left.\left(\frac{\partial D}{\partial \mu}\right)_{U}\right|_{I}}
{\left.\left(\frac{\partial D}{\partial U}\right)_{\mu}\right|_{M}
-\left.\left(\frac{\partial D}{\partial U}\right)_{\mu}\right|_{I}},
\nonumber
\\
&=&
\frac{\left.\left(\frac{\partial D}{\partial \mu}\right)_{U}\right|_{M}}
{\left.\chi_{D}\right|_{M}-\left.\chi_{D}\right|_{I}}, 
\label{eq:Umu1}
\end{eqnarray}
where $(\partial D/\partial \mu)_{U}|_{I}=0$ and eq.~(\ref{eq:doblonsus}) are used. 

On the other hand, we can derive another expression of $\delta U/\delta \mu$ starting from 
the expansion of filling $n$~\cite{WI}. 
In the parallel discussion as above, the slope of the metal-insulator transition boundary 
is derived as 
\begin{eqnarray}
\frac{\delta U}{\delta \mu}= -\frac{\chi_{c}}{\left(\frac{\partial n}{\partial U}\right)_{\mu}|_{M}}. 
\label{eq:Umu2}
\end{eqnarray}
If $(\partial n/\partial \mu)_{U}$ and $(\partial D/\partial U)_{\mu}$ 
are continuous at $(\mu, U)$, 
the following relation holds:
\begin{eqnarray}
\left(\frac{\partial n}{\partial U}\right)_{\mu}
&=& -\frac{1}{N}
\frac{\partial}{\partial U}
\left(\frac{\partial \langle H \rangle}{\partial \mu}
\right)_{U}
\nonumber
\\
&=& -\frac{1}{N}
\frac{\partial}{\partial \mu}
\left(\frac{\partial \langle H \rangle}{\partial U}
\right)_{\mu}
=-\left(\frac{\partial D}{\partial \mu}
\right)_{U}. 
\label{eq:dn_U_dD_mu}
\end{eqnarray}
We then obtain 
\begin{eqnarray}
\left.\left(
\frac{\partial n}{\partial U}
\right)_{\mu}^{2}\right|_{M}
=
\left.\left(
\frac{\partial D}{\partial \mu}
\right)_{U}^{2}\right|_{M}
=
\chi_{c}
(\chi_{D}|_{M}-\chi_{D}|_{I}). 
\label{eq:result}
\end{eqnarray}
By applying this equation to the thermodynamic relations of eq.~(\ref{eq:Umu1}) and eq.~(\ref{eq:Umu2}), 
the following Theorem 1 is derived: 
{\it 
When the metal-insulator transition is a continuous one with a finite and nonzero slope 
in the $\mu$-$U$ phase diagram, 
the divergence of the charge compressibility occurs simultaneously with the divergence of 
the doublon compressibility. 
If the slope of the phase boundary is zero, $\delta U/\delta \mu=0$, 
$\chi_{c} \to \infty \Rightarrow \chi_{D}|_{M}\to \infty$ still holds, 
whereas the converse is not necessarily true. 
}

It should be noted that 
Theorem~1 derived above holds 
irrespective of the spatial dimensionality. 
The conditions in which Theorem~1 holds 
are as follows: 
First, there exists a finite length of the continuous 
metal-insulator-transition boundary $U(\mu)$ in the $\mu$-$U$ plane, 
where $\delta U(\mu)/\delta \mu$ exists. 
Second, the slope of the metal-insulator boundary is 
finite, {\it i.e}., $|\delta U/\delta \mu| < \infty$. 
If the metal-insulator transition takes place as the continuous one 
with the U-shaped structure as shown in Fig.~\ref{fig:UVYshape}(a), 
the thermodynamic relation holds at any point of the transition line. 

\subsection{The case where the slope $\delta U/\delta \mu$ is not well defined} 

\begin{figure}[tb]
\begin{center}
\includegraphics[width=8cm]{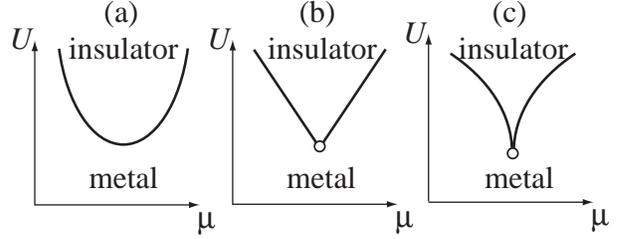}   
\end{center}
\caption{
Classification of the shapes of the phase diagrams 
in the plane of the chemical potential $\mu$ and the Coulomb interaction $U$: 
(a) U shape, (b) V shape, and (c) $\Upsilon$ shape. 
The solid lines represent the continuous metal-insulator transition. 
In (b) and (c), the open circle located at the corner indicates 
the bandwidth-control Mott transition 
point to which the thermodynamic relation cannot be applied (see text). 
}
\label{fig:UVYshape}
\end{figure}

When the Mott-insulator phase has the V-shaped structure 
in the $\mu$-$U$ phase diagram as shown in Fig.~\ref{fig:UVYshape}(b), 
the above discussion is not 
applied at the corner of the V-shaped boundary, which 
is the bandwidth-control Mott-transition point, 
since $\delta U/\delta \mu$ is not well defined at the point. 
The V-shaped Mott-insulator phase in the $\mu$-$U$ plane seems to 
appear actually in the Hubbard model on the square lattice 
for $t'/t=-0.2$ with $t$ and $t'$ being the 
nearest- and next-nearest-neighbor transfers~\cite{WI}. 
In this case, the first-order bandwidth-control Mott transition at the corner 
is compatible with 
the continuous filling-control Mott transition at the edges. 
This offers an example that the corner of the V shape can have the different character of the 
transition from the edges. 

In the one-dimensional Hubbard model without the next-nearest-neighbor hopping, 
the metal-insulator phase boundary has a shape of $\mu \sim \sqrt{tU} {\rm exp}(-2{\pi}t/U)$
for the small-$U/t$ regime~\cite{LiebWu}. 
In the Hubbard model on the square lattice without the next-nearest-neighbor hopping, 
the metal-insulator phase boundary is expected to have 
 a shape of $\mu \sim t {\rm exp}(-2{\pi}\sqrt{t/U})$
for the small-$U/t$ regime~\cite{Hirsch}. 
Both systems with the perfect nesting have the essential-singular form of the 
phase boundary, which is classified into the $\Upsilon$-shaped case 
as shown in Fig.~\ref{fig:UVYshape}(c). 
This is categorized to 
the same class as the V-shaped case.
Namely, the thermodynamic relation can be applied except for 
the corner of the $\Upsilon$-shaped boundary. 

%


\section{Analytic form of doublon susceptibility}

In this section, we derive some useful analytic forms of 
$\partial n/\partial U$, $\partial D/\partial \mu$ and $\chi_{D}=-\partial D/\partial U$ 
near the continuous Mott transition, starting from the analytic forms of $\chi_{c}$ 
realized in the one- and two-dimensional Hubbard models, 
and examine the thermodynamic relations, eqs.~(\ref{eq:Umu1}), (\ref{eq:Umu2}) 
and (\ref{eq:dn_U_dD_mu}) derived in \S2.  
In this section, we approach the transition point from the metallic side.

\subsection{Derivation in one- and two-dimensional Hubbard models}

Let us consider the vicinity of the metal-insulator transition 
in the metallic side of the $\mu$-$U$ phase 
diagram. 
In the one-dimensional Hubbard model without next-nearest-neighbor hopping~\cite{UKO}, 
and in the two-dimensional Hubbard models with~\cite{FI2,WI} and without~\cite{FI} 
the next-nearest-neighbor hoppings, 
the filling dependence of the chemical potential 
is proposed to be scaled as 
\begin{eqnarray}
\mu=\mu_{c}-\frac{1}{2\alpha}\delta^{2}. 
\label{eq:mu-delta}
\end{eqnarray}
with $\delta \equiv 1-n$ and a constant $\alpha$. 
In the one-dimensional system, eq.~(\ref{eq:mu-delta}) was derived 
by the Bethe-Ansatz analysis~\cite{UKO} 
and in the two-dimensional system, eq.~(\ref{eq:mu-delta}) was identified 
by the QMC calculations~\cite{FI,FI2}. 
Note that here we follow the notation of the chemical potential in refs.~\cite{FI,FI2} 
to facilitate the comparison. 
From eq.~(\ref{eq:mu-delta}), 
the doping rate is expressed by the chemical potential as 
\begin{eqnarray}
\delta=\sqrt{2\alpha (\mu_{c}-\mu)}. 
\label{eq:del_mu}
\end{eqnarray}
We need to consider the electron doping separately when 
the next-nearest neighbor hopping 
becomes nonzero causing the electron-hole asymmetry. 
Here, we consider the hole-doped case, 
although the following discussion can easily be extended 
to the electron-doped case. 
Then, the filling dependence of the charge compressibility is given by 
\begin{eqnarray}
\chi_{c}=\frac{\alpha}{1-n}=\frac{\alpha}{\delta}. 
\label{eq:kai_n}
\end{eqnarray}
By substituting eq.~(\ref{eq:del_mu}) to eq.~(\ref{eq:kai_n}), 
the chemical-potential dependence of the charge compressibility 
is given by 
\begin{eqnarray}
\chi_{c}=\sqrt{\frac{\alpha}{2}}(\mu_{c}-\mu)^{-1/2}. 
\label{eq:kai_mu}
\end{eqnarray}

%
To facilitate the discussion below, 
the most divergent terms of 
$\partial D/\partial \mu$, 
$\partial n/\partial U$ and $\chi_{D}=-\partial D/\partial U$ 
near the metal-insulator transition are 
characterized by the exponents $p$, $q$ and $r$ defined below. 

\begin{figure}[tb]
\begin{center}
\includegraphics[width=5cm]{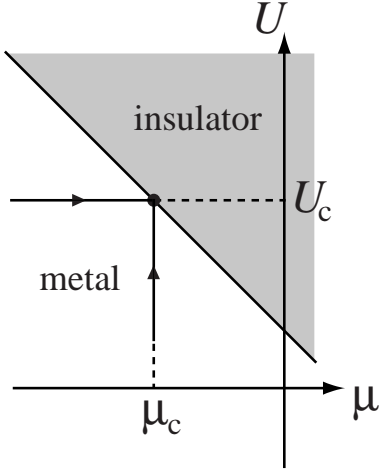}
\end{center}
\caption{
Two ways of approaching the metal-insulator transition point denoted by 
the black circle at $(\mu_{c}, U_{c})$ in the $\mu$-$U$ plane: 
One is $U \to U_{c}-0$ at $\mu=\mu_{c}$ and the other is $\mu \to \mu_{c}-0$ 
at $U=U_{c}$. These are the routes from the metallic side. 
The shaded area is the insulating phase and the solid line is the phase boundary. 
}
\label{fig:limit}
\end{figure}

We consider the metal-insulator phase boundary illustrated in Fig.~\ref{fig:limit}. 
Two routes approaching the transition point at $(\mu_{c}, U_{c})$ 
indicated by the two arrows 
are now the subject of the present study: 
One is $U \to U_{c}-0$ at $\mu=\mu_{c}$ and the other is $\mu \to \mu_{c}-0$ at $U=U_{c}$. 


%

At $U=U_{c}$, 
the relation between $n$ and $\mu$ was already given in eq.~(\ref{eq:del_mu}), 
while the double occupancy 
as a function of $\mu$ near $\mu_{c}$ 
may be similarly assumed as 
\begin{eqnarray}
D= \left\{
\begin{array}{rl}
c(\mu_{c}-\mu)^{1-p}+D_{hf} & \quad \mbox{for $\mu \le \mu_{c}$}, \\
D_{hf}                      & \quad \mbox{for $\mu_{c} \le \mu \le 0$}. 
\end{array}\right. 
\label{eq:D_mu}
\end{eqnarray}
Here $D_{hf}$ is the double occupancy at half filling 
(in the one-dimensional Hubbard model, 
$D_{hf}$ is analytically calculated in ref.~\cite{Shiba}). 

The most dominant term for the filling 
as a function of $U$ near $U_{c}$ 
is assumed in the following form: 
\begin{eqnarray}
n= \left\{
\begin{array}{rl}
-a(U_{c}-U)^{1-q}+1 & \quad \mbox{for $U \le U_{c}$}, \\
1                   & \quad \mbox{for $U_{c} \le U$}. 
\end{array}\right. 
\label{eq:n_U}
\end{eqnarray}
%

At $\mu=\mu_{c}$, the double occupancy 
as a function of $U$ near $U_{c}$ 
may be assumed in the following scaling form: 
\begin{eqnarray}
D= \left\{
\begin{array}{rl}
d(U_{c}-U)^{1-r}+D_{hf}(U_{c}^{+}) & \quad \mbox{for $U \le U_{c}$}, \\
D_{hf}(U)                  & \quad \mbox{for $U_{c} \le U$}. 
\end{array}\right. 
\label{eq:D_U}
\end{eqnarray}

In the metallic phase, the following equations are derived 
from eq.~(\ref{eq:D_mu}) and eq.~(\ref{eq:n_U}), respectively: 
\begin{eqnarray}
\left.
\left(
\frac{\partial D}{\partial \mu}
\right)_{U}
\right|_{M}
&=&-c(1-p)(\mu_{c}-\mu)^{-p},  
\label{eq:dD_mu}
\\
\left.
\left(
\frac{\partial n}{\partial U}
\right)_{U}
\right|_{M}
&=&a(1-q)(U_{c}-U)^{-q}. 
\label{eq:dn_U}
\end{eqnarray}
From eq.~(\ref{eq:D_U}) the doublon susceptibility is written as 
\begin{eqnarray}
\left.
\chi_{D}\right|_{M}&=& 
d(1-r)(U_{c}-U)^{-r}            \quad \mbox{for $U \le U_{c}$}, 
\label{eq:chi_D_U_M}
\\
\left.
\chi_{D}\right|_{I}&=& 
-\frac{\partial D_{hf}(U)}{\partial U}   \quad \mbox{for $U_{c} \le U$}. 
\label{eq:chi_D_U_I}
\end{eqnarray}
%

In the following, 
the coefficients $a$, $c$, and $d$, and 
the critical exponents $p$, $q$ and $r$ are derived analytically. 
Let us first consider eq.~(\ref{eq:D_mu}).
From eq.~(\ref{eq:mu-delta}), the ground-state energy is obtained by integrating $\mu$ 
over $n$: 
\begin{eqnarray}
E(\delta)=E_{hf}-\mu_{c}\delta+\frac{1}{6\alpha}\delta^{3}. 
\end{eqnarray}
From this equation, the double occupancy is obtained as 
\begin{eqnarray}
D=\frac{\partial E}{\partial U}=D_{hf}-
\left(
\frac{\partial \mu}{\partial U}
\right)_{c}
\delta
-
\frac{1}{6 \alpha^{2}}
\left(
\frac{\partial \alpha}{\partial U}
\right)_{c}
\delta^{3}. 
\label{eq:D123}
\end{eqnarray}
Now we consider the vicinity of half filling in the metallic phase and 
expand $D$ up to the first-order in term of $\delta$. 
By substituting eq.~(\ref{eq:del_mu}) to eq.~(\ref{eq:D123}), 
$D$ is expressed by $\mu$: 
\begin{eqnarray}
D=D_{hf}
-
\left(
\frac{\partial \mu}{\partial U}
\right)_{c}
\sqrt{2\alpha}
(\mu_{c}-\mu)^{1/2}, 
\label{eq:Dmu}
\end{eqnarray}
for the most dominant term. 
By differentiating $D$ by $\mu$, we obtain 
\begin{eqnarray}
\frac{\partial D}{\partial \mu}
=
\sqrt{\frac{\alpha}{2}}
\left(
\frac{\partial \mu}{\partial U}
\right)_{c}
(\mu_{c}-\mu)^{-1/2} 
\label{eq:D_mu_2}
\end{eqnarray}
for the most dominant term. 
Comparing eq.~(\ref{eq:D_mu_2}) with eq.~(\ref{eq:dD_mu}), we obtain 
the relation between the coefficients $c$ in eq.~(\ref{eq:D_mu}) 
and $\alpha$: 
\begin{eqnarray}
c=\sqrt{2\alpha}\left(
-\frac{\partial \mu}{\partial U}
\right)_{c}
\label{eq:c_alpha}, 
\end{eqnarray}
and the critical exponent is obtained as $p=1/2$. 

\begin{figure}[tb]
\begin{center}
\includegraphics[width=5cm]{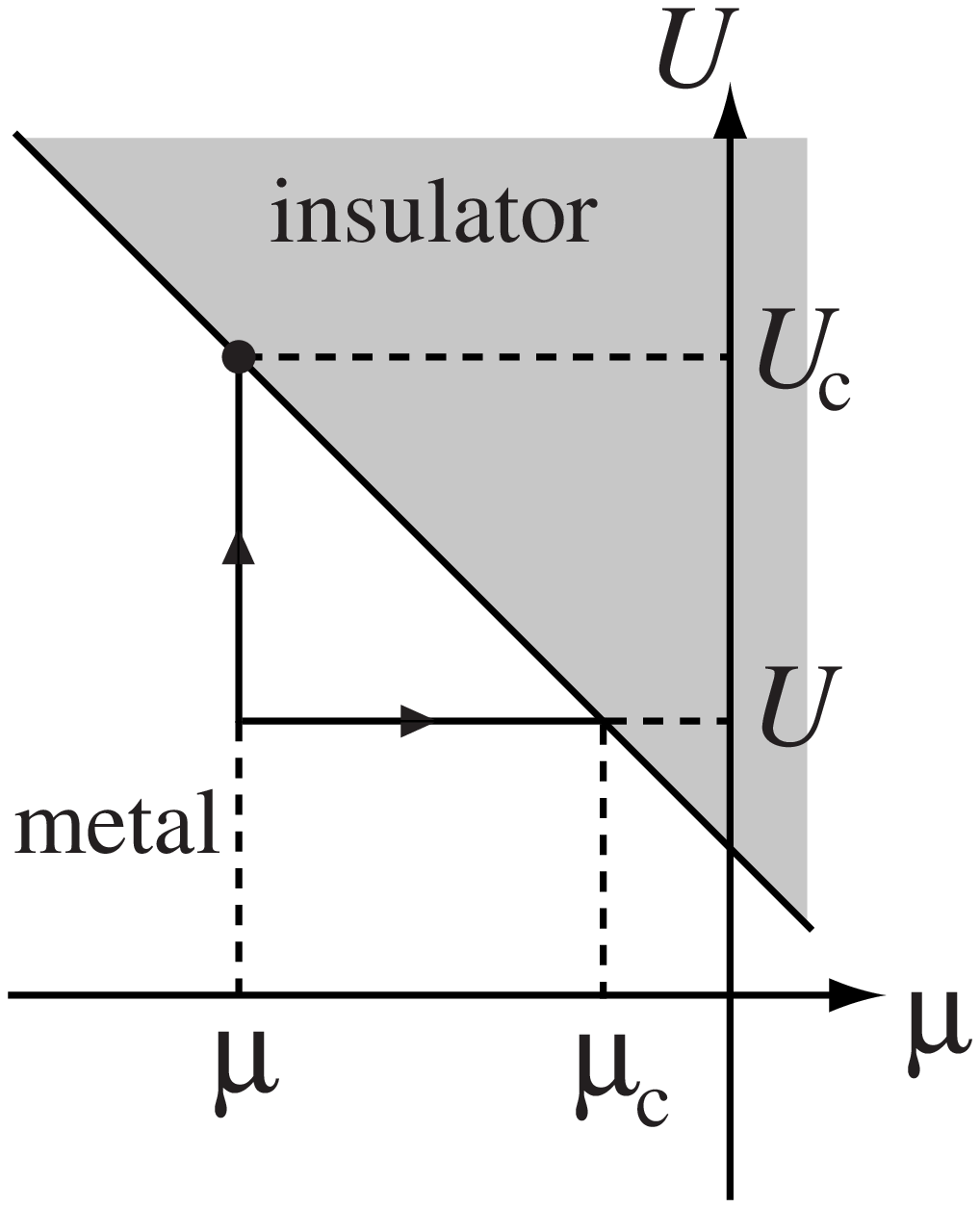}
\end{center}
\caption{
Two ways of approaching the metal-insulator transition point denoted by 
the black circle at $(\mu_{c}, U_{c})$ in the $\mu$-$U$ plane: 
One is $U \to U_{c}-0$ at $\mu=\mu_{c}$ and the other is $\mu \to \mu_{c}-0$ 
at $U=U_{c}$. These are the routes from the metallic side. 
The shaded area is the insulating phase and the solid line is the phase boundary. 
}
\label{fig:limit2}
\end{figure}

Second, let us consider eq.~(\ref{eq:n_U}). 
Starting from eq.~(\ref{eq:del_mu}), we obtain
\begin{eqnarray}
\left(
\frac{\partial \delta}{\partial U}
\right)_{\mu}
&=&\frac{1}{\sqrt{2\alpha}}
\left(
\frac{\partial \alpha}{\partial U}
\right)_{c}
(\mu_{c}-\mu)^{1/2}
\nonumber
\\
&+&\sqrt{\frac{\alpha}{2}}
(\mu_{c}-\mu)^{-1/2}
\left(
\frac{\partial \mu}{\partial U}
\right)_{c},
\label{eq:delta_U_eq}
\end{eqnarray}
%
in the metallic side around $(\mu_{c},U_{c})$. 
Here, we consider the point $(\mu,U)$ close to the phase boundary as in Fig.~\ref{fig:limit2}.
If we define the slope of the phase boundary near $(\mu,U)$ as $(\partial \mu/\partial U)_{c}$, 
the following relation is obviously satisfied: 
%
\begin{eqnarray}
\left.
(\mu_{c}-\mu)
\right|_{U=U_{c}}
=\left(
-\frac{\partial \mu}{\partial U}
\right)_{c}
\left.
(U_{c}-U)
\right|_{\mu=\mu_{c}}
. 
\label{eq:muU}
\end{eqnarray}
In eq.~(\ref{eq:delta_U_eq}), 
in the limit of $\mu \to \mu_{c}-0$, the second term is dominant and 
by using the relation, eq.~(\ref{eq:muU}), 
the diverging part is written as 
\begin{equation}
\frac{\partial n}{\partial U}
=-\frac{\partial \delta}{\partial U}
=\sqrt{\frac{\alpha}{2}}
\left(
-\frac{\partial \mu}{\partial U}
\right)_{c}^{1/2}
(U_{c}-U)^{-1/2}. 
\label{eq:n-U_critical}
\end{equation}
Then, the coefficient $a$ in eq.~(\ref{eq:n_U}) and eq.~(\ref{eq:dn_U}) 
is expressed by $\alpha$ 
and the slope of the metal-insulator boundary:
\begin{equation}
a=\sqrt{2\alpha} \left(
-\frac{\partial \mu}{\partial U}
\right)_{c}^{1/2}, 
\label{eq:a_alpha}
\end{equation}
and the critical exponent is obtained as $q=1/2$. 

%

Third, let us consider eq.~(\ref{eq:D_U}). 
By the derivative of both sides of eq.~(\ref{eq:Dmu}) by $U$, 
the doublon susceptibility is written as
\begin{eqnarray}
\frac{\partial D}{\partial U}
=\frac{\partial D_{hf}}{\partial U}
&-&
\left(
\frac{\partial^{2} \mu}{\partial U^{2}}
\right)_{c}\sqrt{2\alpha}
(\mu_{c}-\mu)^{1/2}
\nonumber
\\
&-&
\left(
\frac{\partial \mu}{\partial U}
\right)_{c}
\frac{1}{\sqrt{2\alpha}}
\left(
\frac{\partial \alpha}{\partial U} 
\right)_{c}
(\mu_{c}-\mu)^{1/2}
\nonumber
\\
&-&
\left(
\frac{\partial \mu}{\partial U}
\right)_{c}^{2}
\sqrt{\frac{\alpha}{2}}
(\mu_{c}-\mu)^{-1/2}. 
\label{eq:DUeq}
\end{eqnarray}
Here $(\partial^{2} \mu/\partial U^{2})_{c}$ is the curvature of the phase boundary. 
In the limit of $\mu \to \mu_c^{-}$, the most dominant term is 
the third term of the right-hand side of eq.~(\ref{eq:DUeq}), 
and by using eq.~(\ref{eq:muU}), we obtain
\begin{eqnarray}
\chi_{D}=
-\frac{\partial D}{\partial U}
=
\sqrt{\frac{\alpha}{2}}
\left(
-\frac{\partial \mu}{\partial U}
\right)_{c}^{3/2}
(U_{c}-U)^{-1/2}. 
\label{eq:chi_D_critial}
\end{eqnarray}
Then, the coefficient $d$ in eq.~(\ref{eq:D_U}) and eq.~(\ref{eq:chi_D_U_M}) 
is expressed by $\alpha$ and the slope of the metal-insulator boundary:
\begin{eqnarray}
d=\sqrt{2\alpha}
\left(
-\frac{\partial \mu}{\partial U}
\right)_{c}^{3/2}, 
\label{eq:d_alpha} 
\end{eqnarray}
and the critical exponent is obtained as $r=1/2$. 


By eqs.~(\ref{eq:c_alpha}), (\ref{eq:a_alpha}) and (\ref{eq:d_alpha})
we have derived the dominant terms of 
$(\partial D/\partial \mu)|_{U=U_{c}}$, 
$(\partial n/\partial U)|_{\mu=\mu_{c}}$ 
and $\chi_{D}=-(\partial D/\partial U)|_{\mu=\mu_{c}}$ 
near the Mott transition, starting from the singularity of $\chi_{c}$. 
Next, let us discuss the relationship between the coefficients of these terms 
and the slope of the metal-insulator boundary, and test Theorem~1.

\subsection{Examination of the thermodynamic relations}

From eqs.~(\ref{eq:c_alpha}), (\ref{eq:a_alpha}) and (\ref{eq:d_alpha}), 
the following relation is derived:
\begin{eqnarray}
\left(
-\frac{\partial U}{\partial \mu}
\right)_{c}
=
\frac{\sqrt{2\alpha}}{c}=\frac{a}{d}=\left(\frac{a}{c}\right)^{2}. 
\label{eq:relation1}
\end{eqnarray}
So far, it is shown that the coefficients 
$a$, $c$ and $d$ are expressed by 
$\alpha$ and $(\partial \mu/\partial U)_{c}$ as in 
eqs.~(\ref{eq:c_alpha}), (\ref{eq:a_alpha}) and (\ref{eq:d_alpha}), 
and the relation of eq.~(\ref{eq:relation1}) is satisfied. 
To examine the thermodynamic relations of eq.~(\ref{eq:Umu1}) and eq.~(\ref{eq:Umu2}), 
let us focus on eq.~(\ref{eq:dn_U_dD_mu}): 
In the metallic phase, $\partial n/\partial U$ and $\partial D/\partial \mu$ 
are continuous and hence eq.~(\ref{eq:dn_U_dD_mu}) holds. 
In the vicinity of the metal-insulator boundary, 
$\partial n/\partial U$ and $\partial D/\partial \mu$ have the form 
of eq.~(\ref{eq:dD_mu}) and eq.~(\ref{eq:dn_U}) and 
the equation holds: 
\begin{eqnarray}
a(U_{c}-U)^{-1/2}=c(\mu-\mu_{c})^{-1/2}. 
\label{eq:a_c_relation}
\end{eqnarray}
By using the relation of eq.~(\ref{eq:muU}), 
the following limit is taken in eq.~(\ref{eq:a_c_relation}): 
\begin{eqnarray}
\lim_{U \to U_{c}^{-}}a(U_{c}-U)^{-1/2}
\nonumber 
\\
=
\lim_{U \to U_{c}^{-}}c
\left(
-
\frac{\partial \mu}{\partial U}
\right)^{-1/2}
(U_{c}-U)^{-1/2}. 
\end{eqnarray}
We see that this equation is satisfied by eq.~(\ref{eq:relation1}). 
This implies that eq.~(\ref{eq:dn_U_dD_mu}) holds in the limit to the metal-insulator boundary 
from the metallic phase in the $\mu$-$U$ plane as it should be 
\begin{eqnarray}
\lim_{U \to U_{c}^{-}} a (U_{c}-U)^{-1/2}=
\lim_{\mu \to \mu_{c}^{-}} c (\mu_{c}-\mu)^{-1/2}.  
\label{eq:aclimit}
\end{eqnarray}
By using eq.~(\ref{eq:aclimit}), eq.~(\ref{eq:relation1}) is 
rewritten as 
\begin{eqnarray}
\left(
-\frac{\partial U}{\partial \mu}
\right)_{c}
&=&
\frac{\lim_{\mu \to \mu_{c}^{-}}\sqrt{2\alpha}(\mu_{c}-\mu)^{-1/2}}
{\lim_{U \to U_{c}^{-}} a(U_{c}-U)^{-1/2}}, 
\nonumber
\\
&=&
\frac{\lim_{\mu \to \mu_{c}^{-}} c(\mu_{c}-\mu)^{-1/2}}
{\lim_{U \to U_{c}^{-}}  d(U_{c}-U)^{-1/2}}. 
\label{eq:relation2}
\end{eqnarray}
This demonstrates how the thermodynamic relations of 
eq.~(\ref{eq:Umu1}) and eq.~(\ref{eq:Umu2}) 
with a simultaneous diverging of the charge compressibility and 
the doublon susceptibility hold.
Then, the validity of the thermodynamic relations is confirmed analytically 
in the case with $\chi_{c}=\sqrt{\alpha/2}(\mu_{c}-\mu)^{-1/2}$ 
which is known to be realized in the one- and two-dimensional Hubbard models.  

\subsection{Bethe-Ansatz solutions in one-dimensional Hubbard model}

\begin{figure}[tb]
\begin{center}
\includegraphics[width=6cm]{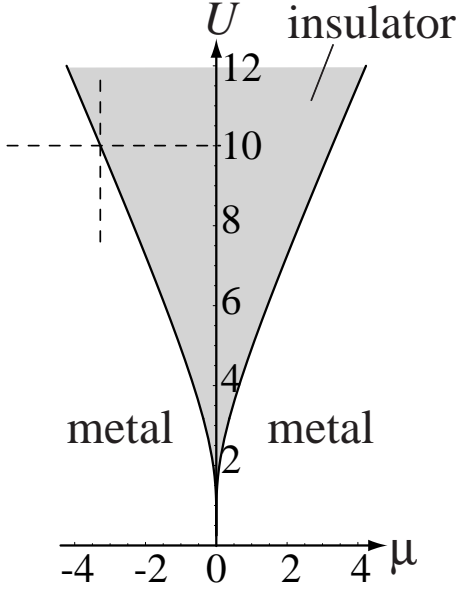}   
\end{center}
\caption{
Ground-state phase diagram of the one-dimensional Hubbard model for $t=1$. 
The metal-insulator phase boundary is obtained by the Bethe-Ansatz solutions~\cite{LiebWu}. 
}
\label{fig:cgap1D}
\end{figure}

In the one-dimensional system, the exact solutions of the Hubbard model 
are available by the Bethe-Ansatz equations~\cite{LiebWu,Shiba,UKO}. 
The energy and the double occupancy at zero temperature for the given 
filling and Coulomb interaction were calculated in ref.~\cite{Shiba}, 
and the coefficient of the charge compressibility $\alpha$ in eq.~(\ref{eq:mu-delta}) 
was obtained in ref.~\cite{UKO}. 
The chemical potential at which the metal-insulator transition occurs for $U$ 
is given by 
\begin{eqnarray}
\mu_{c}^{\pm}=\pm \frac{U}{2} \mp 2t
\left[
1-\int_{0}^{\infty}{\rm d}w\frac{J_{1}(w)}{w(1+{\rm e}^{Uw/2})}
\right], 
\label{eq:mucexact1}
\end{eqnarray}
where upper(lower) sign denotes the electron(hole)-doped regime and 
$J_{1}(w)$ is the first Bessel function~\cite{LiebWu}. 
The ground-state phase diagram of the one-dimensional Hubbard model 
is shown in Fig.~\ref{fig:cgap1D}. 
By using the exact expressions of $D$ in eqs.~(2.14) and (2.15) in ref.~\cite{Shiba}
and $n$ in eqs.~(6a) and (6b) in ref.~\cite{UKO}, we plot 
the critical behaviors of these quantities as functions of $\mu$ 
and $U$ near the Mott transition. 
As a typical example, we show the critical behaviors of $n$ and $D$ 
around the Mott transition point, $(\mu_{c},U_{c})=(-3.277,10.0)$ 
(see Fig.~\ref{fig:cgap1D}).

\begin{figure}[tb]
\begin{center}
\includegraphics[width=7cm]{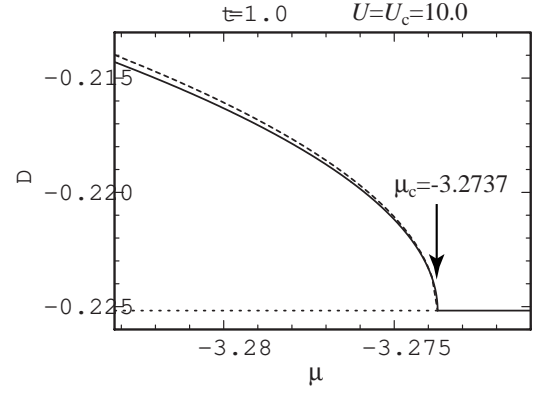}   
\end{center}
\caption{
Double occupancy $D$ vs. chemical potential $\mu$ obtained by the Bethe-Ansatz solutions 
of the one-dimensional Hubbard model (solid line) 
for $t=1$ at $U=U_{c}=10$.  
Dashed line represents 
$D=D_{hf}+\sqrt{2\alpha}(-\partial \mu/\partial U)_{c}(\mu_{c}-\mu)^{-1/2}$. 
}
\label{fig:D-mu}
\end{figure}

The chemical-potential dependence of the double occupancy for $U=U_{c}$
is shown in Fig.~\ref{fig:D-mu}. 
We also plot 
$D=D_{hf}+\sqrt{2\alpha}(-\partial \mu/\partial U)_{c}(\mu_{c}-\mu)^{-1/2}$ 
obtained from eqs.~(\ref{eq:Dmu}) and (\ref{eq:c_alpha}) 
by the dashed line. 
We see that $\partial D/\partial \mu$ diverges at $\mu=\mu_{c}$, 
whose critical behavior is described by eq.~(\ref{eq:D_mu_2}).

%

\begin{figure}[tb]
\begin{center}
\includegraphics[width=7cm]{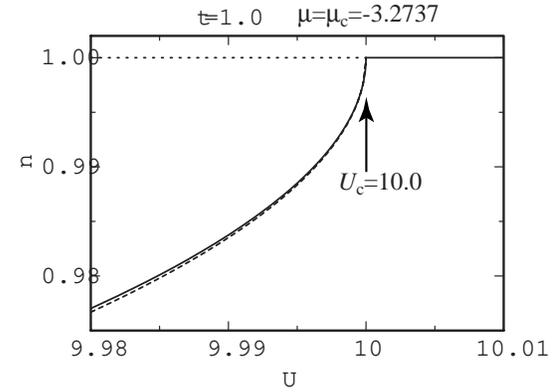}   
\end{center}
\caption{
Filling $n$ vs. Coulomb interaction $U$ 
obtained by the Bethe-Ansatz solutions 
of the one-dimensional Hubbard model
for $t=1$ at $\mu=\mu_{c}=-3.2737$. 
Dashed line represents 
$n=1-\sqrt{2\alpha}(-\partial \mu/\partial U)_{c}^{1/2}(U_{c}-U)^{1/2}$. 
}
\label{fig:n-U}
\end{figure}

The interaction dependence of the filling for $\mu=\mu_{c}$ 
is shown in Fig.~\ref{fig:n-U}. 
We also plot 
$n=1-\sqrt{2\alpha}(-\partial \mu/\partial U)_{c}^{1/2}(U_{c}-U)^{1/2}$ 
obtained from eqs.~(\ref{eq:n_U}) and (\ref{eq:a_alpha}) 
by the dashed line. 
We see that $\partial n/\partial U$ diverges at $U=U_{c}$, whose 
critical behavior is described by eq.~(\ref{eq:n-U_critical}).

%

%

\begin{figure}[tb]
\begin{center}
\includegraphics[width=7cm]{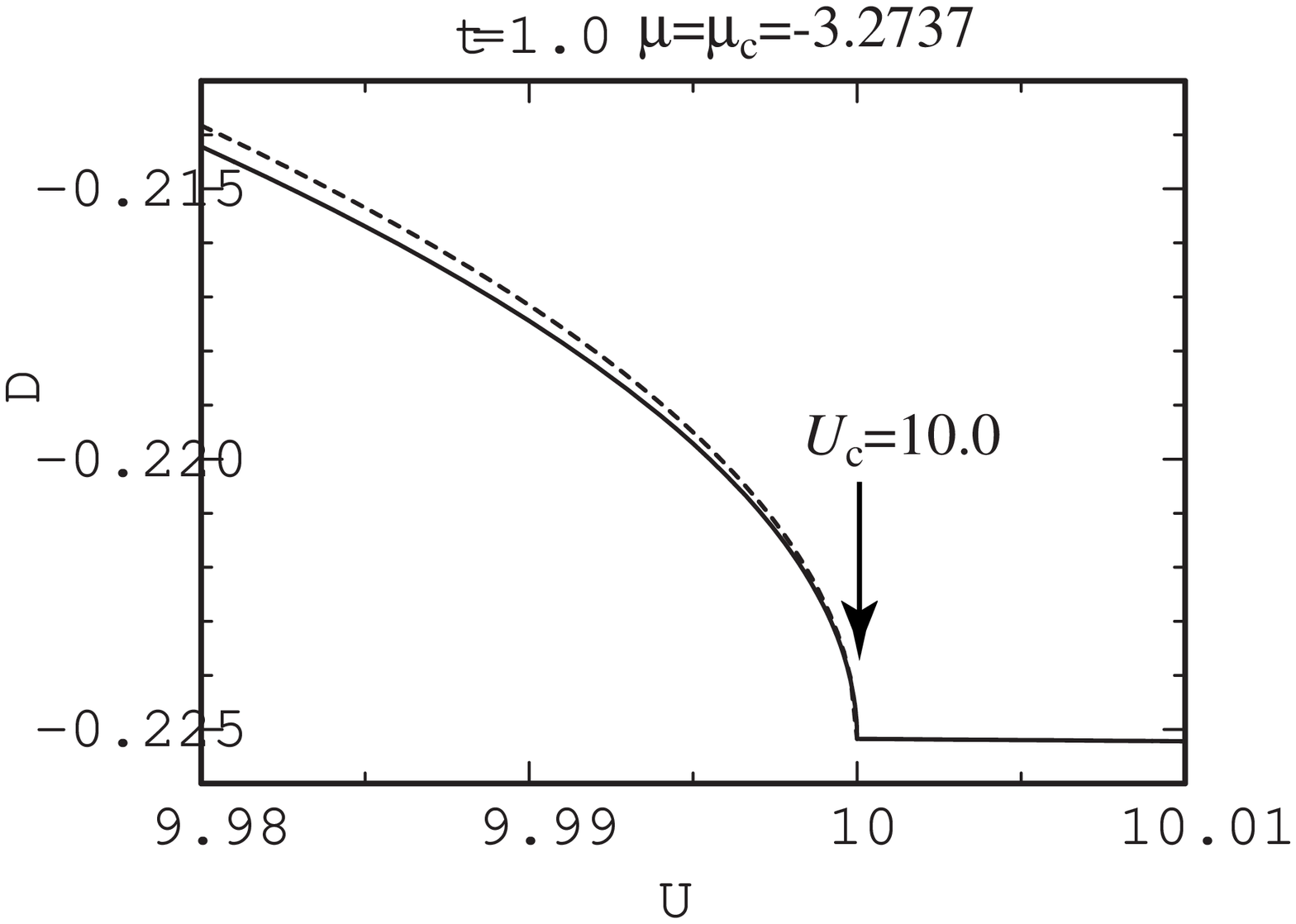}   
\end{center}
\caption{
Double occupancy $D$ vs. Coulomb interaction $U$ 
obtained by the Bethe-Ansatz solutions 
of the one-dimensional Hubbard model
for $t=1$ at $\mu=\mu_{c}=-3.2737$.    
Dashed line represents 
$D=D_{hf}+\sqrt{2\alpha}(-\partial \mu/\partial U)_{c}^{3/2}(U_{c}-U)^{1/2}$. 
}
\label{fig:D-u}
\end{figure}

The interaction dependence of the double occupancy for $\mu=\mu_{c}$ 
is shown in Fig.~\ref{fig:D-u}. 
We also plot 
$D=D_{hf}+\sqrt{2\alpha}(-\partial \mu/\partial U)_{c}^{3/2}(U_{c}-U)^{1/2}$ 
obtained from eqs.~(\ref{eq:D_U}) and (\ref{eq:d_alpha}) 
by the dashed line. 
We see that the doublon susceptibility diverges at $U=U_{c}$ 
and its critical behavior is described by 
eq.~(\ref{eq:chi_D_critial}). 

%


In the one-dimensional Hubbard model, the limits of $\alpha$ 
in eq.~(\ref{eq:mu-delta}) were shown as $\alpha \to 0$ for $U \to 0$ 
and $\alpha \to 1/(2\pi^{2})$ for $U \to \infty$~\cite{UKO}. 
Since $(\partial \mu/\partial U)_{c} \to 0$ for $U \to 0$ 
and $(\partial \mu/\partial U)_{c} \to 1/2$ for $U \to \infty$, 
the limits of $a$ ,$c$ and $d$ are as follows: 
$a \to 0$, $c \to 0$, and $d \to 0$ for $U \to 0$, and 
$a \to 1/(\sqrt{2}\pi)$, $c \to 1/(2\pi)$ 
and $d \to 1/(2\sqrt{2}\pi)$ for $U \to \infty$. 

\subsection{General expressions in $d$-dimensional systems with dynamical exponents $z$}

We have discussed the analytic forms of 
$\chi_{c}|_{U=U_{c}}$, $(\partial n/\partial U)|_{\mu=\mu_{c}}$, 
$(\partial D/\partial \mu)|_{U=U_{c}}$ and $\chi_{D}|_{\mu=\mu_{c}}$, 
which are realized in the one- and two-dimensional Hubbard models. 
Here we present the general expressions of these quantities in $d$-dimensional systems. 

Based on the hyperscaling analysis, the critical behavior of physical quantities 
near the continuous filling-control Mott transition was predicted in ref.~\cite{Imada}
For example, the charge compressibility has the form of 
\begin{eqnarray}
\chi_{c}=\sqrt{\frac{\alpha}{2}}(\mu_{c}-\mu)^{\nu (d+z)-2}, 
\label{eq:chi_general}
\end{eqnarray}
where $\nu$ is the correlation-length exponent which satisfies $\nu z=1$ with 
$z$ being the dynamical exponent, and $d$ is the spatial dimension. 
It is known that for the usual one-dimensional Hubbard model, where 
$d=1$, the dynamical exponent 
is given by $z=2$, then the charge compressibility has the form of 
$\chi_{c} \sim (\mu_{c}-\mu)^{-1/2}$ as in eq.~(\ref{eq:kai_mu}). 
For the two-dimensional system, $d=2$, the dynamical exponent $z=4$ 
was identified by numerical calculations~\cite{AI1,AI2}, and 
then the same critical behavior for $\chi_{c}$ as the $d=1$ case appears. 
For the three-dimensional system, in the case of $z=4$, 
the charge compressibility has the form of $\chi_{c}\sim (\mu_{c}-\mu)^{-1/4}$ 
by eq.~(\ref{eq:chi_general}). 
In the following, we show 
the analytic forms of the physical quantities defined by the second derivative 
of the Hamiltonian in general form, which are derived from eq.~(\ref{eq:chi_general}):
\begin{eqnarray}
\left(
\frac{\partial n}{\partial U}
\right)_{\mu=\mu_{c}}
&=&
\sqrt{\frac{\alpha}{2}}
\left(
\frac{\partial \mu}{\partial U}
\right)_{c}^{\nu (d+z)-1}
\nonumber 
\\
& &
\times
(U_{c}-U)^{\nu (d+z)-2},
\nonumber
\\
\left(
\frac{\partial D}{\partial \mu}
\right)_{U=U_{c}}
&=&
\sqrt{\frac{\alpha}{2}}
\left(
\frac{\partial \mu}{\partial U}
\right)_{c}
\nonumber
\\
& &
\times
(\mu_{c}-\mu)^{\nu (d+z)-2},
\nonumber
\\
\chi_{D}=
-
\left(
\frac{\partial D}{\partial U}
\right)_{\mu=\mu_{c}}
&=&
\sqrt{\frac{\alpha}{2}}
\left(
\frac{\partial \mu}{\partial U}
\right)_{c}^{\nu (d+z)}
\nonumber
\\
& &
\times
(U_{c}-U)^{\nu (d+z)-2}. 
\nonumber
\end{eqnarray}
The coefficients of each term are expressed by the coefficients of $\chi_{c}$ and the slope of 
the metal-insulator boundary. 
The exponents are the same as that of the charge compressibility, {\it i.e}., $\nu (d+z)-2$. 
Note that the relation eq.~(\ref{eq:relation1}) still holds in this case, and 
this implies that eq.~(\ref{eq:relation1}) does not depend on the exponent in the 
divergence of the charge compressibility. 
This is ascribed to the fact that thermodynamic relations of eq.~(\ref{eq:Umu1}) and 
eq.~(\ref{eq:Umu2}) hold in any spatial dimensions. 

\section{Finite temperature}
\subsection{Generalization of thermodynamic relations to finite temperature}
In \S3, the thermodynamic relations are examined at zero temperature. 
It should be noted, however, that the thermodynamic relations of eqs.~(\ref{eq:derivum}), 
(\ref{eq:Umu1}) and (\ref{eq:dn_U_dD_mu}) hold even at finite temperature,  
if $\langle ... \rangle$ is taken as the thermal average:
\begin{eqnarray}
\langle A \rangle \equiv 
\frac{{\rm tr}\left({\rm e}^{-H/T} A \right)}{{\rm tr}\left({\rm e}^{-H/T}\right)}, 
\label{eq:thermal_av}
\end{eqnarray}
where $A$ is an operator for a physical quantity and $T$ is temperature. 
Hence, when temperature is fixed, in the plane of $\mu$ and $U$ with the finite length 
of the metal-insulator boundary, 
eqs.~(\ref{eq:derivum}), (\ref{eq:Umu1}) and (\ref{eq:dn_U_dD_mu}) hold. 

\subsection{Constraints on  the shape of the phase diagram for finite-temperature 
first-order metal-insulator transition}
We also discuss the phase diagram with the temperature dependence. 
Starting from the expansion of the Gibbs free energy $G(T,U)$ for electrons 
along the first-order 
metal-insulator-transition line, 
we obtain
\begin{eqnarray}
\frac{\delta U}{\delta T}=\frac{S_{I}-S_{M}}{D_{I}-D_{M}}, 
\label{eq:dUdT}
\end{eqnarray}
where the entropy of electrons is defined by $S \equiv -(\partial G/\partial T)_{U}$ 
and $D$ is the double occupancy in the thermal average.
In experiments, when the pressure $p$ is applied to the system, 
the bandwidth of the electronic system, $U/t$ changes. 
Hence, in eq.~(\ref{eq:dUdT}), $U$ corresponds to $p$ and the double occupancy 
$D$ is related to the conjugate variable, the volume $v$. 
The similarity between the metal-insulator transition and the liquid-gas transition 
was pointed out in ref.~\cite{castellani}. 

\begin{figure}[tb]
\begin{center}
\includegraphics[width=7.5cm]{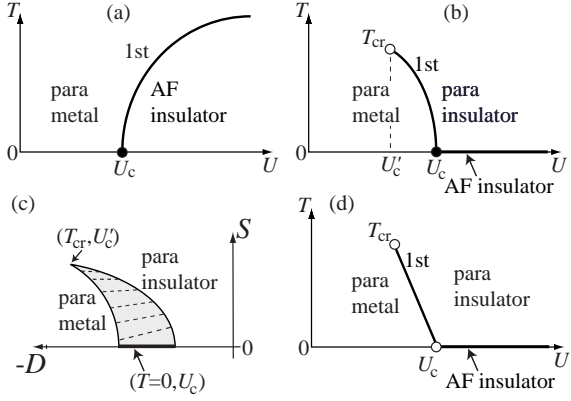}
\end{center}
\caption{
Some examples of 
schematic phase diagrams with first-order metal-insulator transition 
in the plane of temperature $T$ and 
Coulomb interaction $U$. 
(a): Paramagnetic metal and anfiferromagnetic insulator phases 
in the three dimensional system. 
(b): Paramagnetic metal phase for $T \ge 0$, the anfiferromagnetic insulator phase at $T=0$ and 
the paramagnetic insulator phase for $T>0$ in the two dimensional system. 
Open circle in (b) represents critical end point, $T_{cr}$, of the first-order 
transition line. 
(c): Phase diagram in the plane of double occupancy $D$ and entropy $S$ corresponding to (b). 
Gray area represents the phase-separated region between metal and insulator 
along the first-order-transition line in (b). 
Dashed line represents each set of $(T,U)$ on the first-order-transition line in (b). 
(d): Schematic phase diagram for the continuous transitions both at $T=0$ and $T=T_{cr}$ 
drawn by open circles, 
while the first-order transition occurs for $0 < T < T_{cr}$. 
In this case, a finite slope 
$(\partial T/\partial U)|_{T \to 0} \ne \infty$ 
is allowed in contrast to (a) and (b)(see text). 
}
\label{fig:UT}
\end{figure}

We discuss the consequence of eq.~(\ref{eq:dUdT}). 
Let us consider the case that the first-order metal-insulator transition takes place in the 
$U$-$T$ phase diagram and the insulator phase has the ordering such as the magnetic one. 
In the metallic phase, 
at small but nonzero 
particle-hole excitations induce the entropy stemming from the finite 
density of states at the Fermi level, 
while in the insulator phase the entropy is rather small 
if the symmetry breaking with small degeneracy 
is realized while the particle-hole excitation 
cannot induce the entropy under the finite charge gap. 
Then, this yields $S_{I}-S_{M}<0$ and by using the fact that $D_{I}-D_{M}<0$, 
the slope should be $\delta T/\delta U >0$ (see Fig.~\ref{fig:UT}(a)). 
This is a reasonable result, 
since the metallic phase appears in the high-temperature side to govern the 
$-TS$ term in the free energy. 
This situation can be realized in the three dimensional system, 
since Mermin-Wagner's theorem~\cite{MW} allows the magnetic ordering at finite temperature. 

In the two-dimensional system, the magnetic ordering at finite temperature 
is suppressed by thermal fluctuation~\cite{MW}. 
In this case, the paramagnetic insulator phase can be realized for finite-temperature regime 
above the magnetically-ordered insulator phase at $T=0$. 
Although 
it depends on each system 
how the entropy is released by the inter-site interaction 
in the paramagnetic-insulator phase, 
we refer ref.~\cite{OI} as an example of the appearance of the paramagnetic insulator phase 
at high-temperature side, {\it i.e}., $\delta T/\delta U <0$ by $S_{I}-S_{M}>0$ 
(see Fig.~\ref{fig:UT}(b)). 

%

Let us focus on the slope of the first-order transition in the zero-temperature limit; 
$(\partial T/\partial U)|_{T \to 0}$. 
When the first-order metal-insulator transition occurs at $T=0$, the jump of 
the double occupancy appears at $U=U_{c}$, {\it i.e}., $D_{I}-D_{M} \ne 0$. 
By the requirement of the thermodynamic third law, the entropy at $T=0$ should be 0 
in the metal and insulator phases, 
and hence $S_{I}=0=S_{M}$ at $T=0$. 
Therefore, from eq.~(\ref{eq:dUdT}), 
the following conclusion is derived: 
{\it
In the phase diagram with temperature $T$ and control parameter $U$, 
the first-order transition line should have the infinite slope 
in the zero-temperature limit: 
$(\partial T/\partial U)|_{T \to 0}=\infty$. 
}
This statement is not restricted to the case of the metal-insulator transition, 
but can be applied to the first-order transition between any types of the phases. 
The schematic phase diagram in the plane of the double occupancy $D$ and entropy $S$  
corresponding to Fig.~\ref{fig:UT}(b) is illustrated in Fig.~\ref{fig:UT}(c). 
The gray area represents the phase-separated region between the metal and the insulator 
along the finite-temperature first-order transition line in Fig.~\ref{fig:UT}(b). 
The dashed line in  Fig.~\ref{fig:UT}(c) indicates each set of $(T,U)$ 
on the finite-temperature first-order transition line in Fig.~\ref{fig:UT}(b).

Here, the relationship to the known results should be mentioned. 
In the infinite-dimensional Hubbard model with the fully frustrated lattice, 
the $T$-$U$ phase diagram was obtained~\cite{GK,RKZ}: 
The first-order transition between the paramagnetic metal and the paramagnetic insulator 
takes place at finite temperature. 
One might consider that the first-order-transition line has a finite slope as 
$(\partial T/\partial U)|_{T \to 0}<0$ in the phase diagram, 
which is different from the above statement. 
However, at $T=0$ the metal-insulator transition takes place as the continuous one 
in that model, to which the above discussion cannot be applied. 
Actually, in this case, the continuous transition with $D_{I}-D_{M}=0$ occurs at $T=0$. 
Then, the finite slope is possible; $|\partial T/\partial U|_{T \to 0} < \infty$. 
The schematic phase diagram in the $U$-$T$ plane for such a case 
is illustrated in Fig.~\ref{fig:UT}(d):  
At $T=0$ and $T=T_{cr}$, the continuous metal-insulator transitions take place, while 
the first-order transition occurs for $0 < T < T_{cr}$.

At $T=0$, possible shapes of the phase diagram with the first-order and 
continuous quantum phase transitions have been classified in ref.~\cite{WI}. 
In the plane of the chemical potential $\mu$ and the Coulomb interaction $U$, 
the U-shaped phase boundary with a finite length of the first-order transition can exist, 
but the V-shaped phase is allowed to have the only first-order-transition point at the corner 
and have the continuous transitions at the edges. 
The $\Upsilon$-shaped phase as an essential singular form of the 
Mott gap, $\sim {\rm exp}[-\alpha t/(U_{c}-U)]$ is classified into the same class as the V-shaped case.

\begin{figure}
\begin{center}
\includegraphics[width=7.5cm]{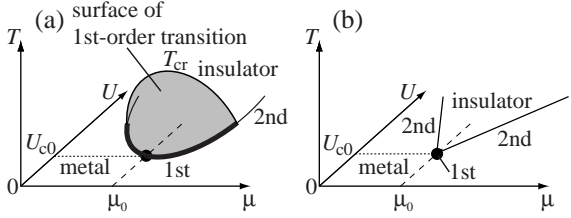}
\end{center}
\caption{
Possible schematic phase diagrams in the parameter space 
of temperature $T$, chemical potential $\mu$ 
and interaction $U$ for (a) U-shaped and (b) V-shaped Mott-insulator phases at $T=0$. 
In (a), thick(thin) line represents the first-order(continuous) transition line. 
Gray surface represents the surface of the first-order transition and its upper edge is 
the critical temperature, $T_{cr}$. 
In (a) and (b), 
the solid circle represents the band-width control Mott transition point at $U_{c0}$, 
and $\mu_{0}$ represents the chemical potential at half filling at $U=U_{c0}$. 
The dashed line connected to $\mu_{0}$ is the route of the band-width control. 
In (b), the first-order transition takes place only at the bandwidth-control 
Mott-transition point, {\it i.e}., $T_{cr}=0$. 
}
\label{fig:TUmu}
\end{figure}

For the U-shaped case at $T=0$, a possible $T$-$U$-$\mu$ phase diagram is illustrated 
in Fig.~\ref{fig:TUmu}(a). 
The surface of the first-order transition is formed as drawn by the gray area 
in Fig.~\ref{fig:TUmu}(a) and at the critical end curve drawn by the solid line, 
which describes $T_{cr}$, 
the charge compressibility diverges~\cite{Kotliar}. 
For the V-shaped case at $T=0$, the possible $T$-$U$-$\mu$ phase diagram is illustrated 
in Fig.~\ref{fig:TUmu}(b). 
In this case, at $T=0$ the first-order transition occurs at the corner of the V-shaped boundary. 
At finite temperatures, 
one possibility is that the jump of the double occupancy at $T=0$ 
disappears for infinitesimal temperature. 
Actually one can consider such a function for $D(\mu_{0},U,T)$ with $\mu_{0}$ 
the chemical potential at half filling: 
An example is 
the functional form analogous to 
the Fermi distribution function, 
$f(\varepsilon,T) \equiv 1/[\exp((\varepsilon-\mu)/T)+1]$. 
At $T=0$ the jump of $f(\varepsilon,T)$ 
at $\varepsilon=\mu$ appears, 
while at infinitesimal $T$, the jump disappears. 
This is the case illustrated in Fig.~\ref{fig:TUmu}(b) where $T_{cr}$ becomes zero temperature, 
{\it i.e.}, the quantum tricritical point. 
In this case, $D(\mu_{0},U,T=0)$ as a function of $U$ 
has a jump at the bandwidth-control Mott transition point, $U_{c0}$, 
and the jump disappears at $U_{c0}$ for $T \ne 0$. 

The other possibility is that the jump 
in $D(\mu_{0},U,T)$ remains even for $T \ne 0$. 
In this case, the first-order-transition line, but not the surface as in Fig.~\ref{fig:TUmu}(a), 
has a certain length from $T=0$ to $T=T_{cr}$. 
This seems to be unusual, but such a case is not ruled out by the thermodynamic argument.

%

\section{Application of thermodynamic relation to general models}
In {\S}2, we have derived the thermodynamic relations for the metal-insulator 
transition, and in \S3 we have examined that 
physical quantities defined by the second derivative of the Hamiltonian for $\mu$ and $U$ 
actually satisfy the thermodynamic relations 
in the cases of the one- and two-dimensional Hubbard models. 
However, the thermodynamic relation between the slope of the phase boundary and 
physical quantities defined by the first and second derivative of the Hamiltonian 
for the first-order and continuous transitions, respectively, can be derived in general 
in the other models with any types of phases in the same way as in \S2. 

As an example, let us consider the phase diagram whose control parameters are the 
on-site Coulomb interaction $U$ and the nearest-neighbor Coulomb interaction $V$. 
The Hamiltonian is described by the extended form as 
\begin{eqnarray}
H_{ex}=H+
V \sum_{\langle i,j \rangle}\sum_{\sigma\sigma'}
n_{i\sigma}n_{j\sigma'}, 
\label{eq:Hex}
\end{eqnarray}
where $\langle i,j \rangle$ denotes the nearest-neighbor sites. 

At quarter filling, the existence of two phases, namely, metallic and charge-ordering phases 
has been reported in the one-dimensional system~\cite{CO1D}. 

At the first-order transition, by expanding the ground-state energy 
by $U$ and $V$ in a similar way to the derivation of eq.~(\ref{eq:derivum}), 
the slope of the phase boundary 
is expressed as 
\begin{eqnarray}
\frac{\delta U}{\delta V}=-\frac{R|_{M}-R|_{I}}{D|_{M}-D|_{I}}, 
\label{eq:UV1}
\end{eqnarray}
where 
\begin{eqnarray}
R \equiv 
\frac{1}{N}
\frac{\partial \langle H \rangle}{\partial V}
=
\frac{1}{N}
\sum_{\langle i,j \rangle}
\sum_{{\sigma}{\sigma}'}\langle
n_{i\sigma}n_{j\sigma'}
\rangle. 
\nonumber
\end{eqnarray}
Here, at the first-order transition, the slope of the phase boundary in the $U$-$V$ plane 
is expressed by the ratio of the jump of the nearest-neighbor correlation and the double occupancy. 

At the continuous transition, the slope of the phase boundary 
is expressed in two ways: 
One is the expansion of the double occupancy and the other is the 
nearest-neighbor correlation. 
The resultant thermodynamic relations are 
\begin{eqnarray}
\frac{\delta V}{\delta U}&=&
\frac
{
 \left.\chi_{D}\right|_{M}
-\left.\chi_{D}\right|_{I}
}{\left.\left(\frac{\partial D}{\partial V}\right)_{U}\right|_{M}
      -\left.\left(\frac{\partial D}{\partial V}\right)_{U}\right|_{I}}, 
\nonumber \\
&=& \frac{\left.\left(\frac{\partial R}{\partial U}\right)_{U}\right|_{M}
-\left.\left(\frac{\partial R}{\partial U}\right)_{U}\right|_{I}
}
{\left.\chi_{V}\right|_{M}
-\left.\chi_{V}\right|_{I}
}, 
\label{eq:UV2}
\end{eqnarray}
where the susceptibility for the nearest-neighbor correlation 
is defined by 
\begin{eqnarray}
\chi_{V} \equiv 
-\frac{\partial^{2} \langle H \rangle}{\partial V^{2}}. 
\end{eqnarray}
If $(\partial D/\partial V)_{U}$ and $(\partial R/\partial U)_{V}$ are continuous 
at $(U,V)$, the following relation holds, which is shown in the same way  
as in the derivation of eq.~(\ref{eq:dn_U_dD_mu}): 
\begin{eqnarray}
\left(
\frac{\partial D}{\partial V}
\right)_{U}
=
\left(
\frac{\partial R}{\partial U}
\right)_{V}. 
\end{eqnarray}
When we add the axis of the chemical potential to the phase diagram in the plane of $U$ and $V$, 
the thermodynamic relations in the $\mu$-$U$ plane and the $\mu$-$V$ plane 
can be derived as in a parallel way to \S2. 

Here, we show the thermodynamic relations for the first-order and continuous transitions 
in the extended Hubbard model as an example. 
It should be noted that 
the thermodynamic relations can also be applied to other systems in a similar way. 
Hence, we remark that the thermodynamic relations derived in this paper 
have generality. 
The applicability of the relations 
is not restricted to the specific model, but 
to any types of the models with the phase diagram. 
This is a consequence of the extensions of the Clausius-Clapeyron and Ehrenfest relations 
to microscopic models.

\section{Summary}
We have analytically derived several useful 
thermodynamic relations of the charge compressibility and the doublon susceptibility 
to the slope of the metal-insulator phase boundary in the 
plane of the chemical potential $\mu$ and the Coulomb interaction $U$. 
By using this relation, 
we have shown that the charge compressibility divergence at the metal-insulator transition 
causes simultaneous divergence of 
the doublon susceptibility 
when the metal-insulator transition takes place as a continuous one with 
a finite length and a finite and nozero slope in the $\mu$-$U$ plane. 
This statement holds at any spatial dimensions and at any temperatures. 
The main results of the obtained relations are eqs.~(\ref{eq:Umu1}), 
(\ref{eq:Umu2}) and (\ref{eq:result}). 

In one- and two-dimensional Hubbard models, 
critical divergence of the charge compressibility at the Mott transition 
appears generally with the form of 
$\chi_{c}=\sqrt{\frac{\alpha}{2}}(\mu_{c}-\mu)^{-1/2}$. 
Starting from this critically divergent form of $\chi_{c}$, 
analytical expressions of 
$(\partial D/\partial \mu)_{U=U_{c}}$, 
$(\partial n/\partial U)_{\mu=\mu_{c}}$ and $\chi_{D}|_{\mu=\mu_{c}}$ 
in the vicinity of the Mott transition are derived. 
All the coefficients of the most divergent terms of these quantities are expressed by the 
coefficient of the charge compressibility, $\alpha$, and 
the slope of the metal-insulator boundary, $(\delta U/\delta \mu)_{c}$. 
All the exponents of the divergence of these quantities have 
the same value as the exponent of $\chi_{c}$: 
namely, $-1/2$. 
The relations among 
the coefficients of $\chi_{c}|_{U=U_c}$, $(\partial D/\partial \mu)_{U=U_{c}}$, 
$(\partial n/\partial U)_{\mu=\mu_{c}}$ and $\chi_{D}|_{\mu=\mu_{c}}$ 
at the metal-insulator-transition point actually satisfy 
the thermodynamic relations derived above, {\it i.e}., 
$
\left.\left(
\frac{\partial n}{\partial U}
\right)^{2}\right|_{M}
=
\left.\left(
\frac{\partial D}{\partial \mu}
\right)^{2}\right|_{M}
=
\chi_{c}
(\chi_{D}|_{M}-\chi_{D}|_{I})
$. 

The above statement holds at any points of the U-shaped metal-insulator boundary and 
at the points of the V $(\Upsilon)$-shaped boundary except the corner, which is the 
bandwidth-control Mott transition point.  
At the corner of V- and $\Upsilon$-shaped phases $(\delta U/\delta \mu)_{c}$ is not well defined. 

By applying the above argument to the phase diagram 
in the parameter space of 
the temperature $T$ and the interaction $U$, 
constraints on the phase boundary are classified 
at finite temperatures. 
It is shown that the infinite slope of the first-order-transition line
has the infinite slope as 
$(\partial T/\partial U)|_{T \to 0}=\infty$ 
from the requirement of the thermodynamic third law and thermodynamic relations.

These conclusions described above are not restricted only to the Hubbard model, but also 
can be applied to other general models. 
As an example, we demonstrate the derivation of the thermodynamic relations 
in the extended Hubbard model with the next-near-neighbor repulsion.

\section*{Acknowledgment}
One of the authors (S. W.) would like to thank Prof. Minoru Takahashi, Prof. Masao Ogata 
and Dr. Masahiro Shiroishi for helpful suggestions about the Bethe-Ansatz solutions of 
the one-dimensional Hubbard model. 
The work is supported by Grant-in-Aid for young scientists, No. 15740203 
as well as Grant-in-Aid No.~16340100 
from the Ministry of Education, Culture, Sports, Science and Technology, 
Japan. 
A part of our computation has been done at the Supercomputer Center 
in the Institute for Solid State Physics, University of Tokyo. 



\end{document}